\documentclass[conference]{IEEEtran}
\IEEEoverridecommandlockouts
\usepackage{cite}
\usepackage{amsmath,amssymb,amsfonts}
\usepackage{tikz}
\usepackage{bm}
\usepackage{booktabs}
\usepackage{threeparttable}
\usepackage{algorithm}
\usepackage[noend]{algpseudocode}

\usepackage{hyperref}
\hypersetup{draft}

\usepackage{graphicx}
\usepackage{textcomp}
\usepackage{xcolor}

\def\BibTeX{{\rm B\kern-.05em{\sc i\kern-.025em b}\kern-.08em
    T\kern-.1667em\lower.7ex\hbox{E}\kern-.125emX}}
    
\usepackage{xcolor}
\definecolor{difftitle}{HTML}{000099}
\definecolor{diffstart}{HTML}{660099}
\definecolor{diffincl}{HTML}{006600}
\definecolor{diffrem}{HTML}{AA3300}

\usepackage{listings}
\usepackage{booktabs}
\usepackage{multirow}
\usepackage[justification=centering]{caption}
\usepackage{subfig}
\usepackage{tabularx,booktabs}

\lstdefinelanguage{diff}{
    backgroundcolor=\color{white},  % choose the background color
    basicstyle=\ttfamily\small,
    morecomment=[f][\color{difftitle}]{diff},
    morecomment=[f][\color{difftitle}]{index},
    morecomment=[f][\color{diffstart}]{@@},
    morecomment=[f][\color{diffincl}]{+},
    morecomment=[f][\color{diffrem}]{-},
    columns=fullflexible,
    tabsize=4,
    breaklines=true,% automatic line breaking only at whitespace
    captionpos=b, % sets the caption-position to bottom
    frame=none,
}

\newcommand{\TN}{PatchRNN}

\begin{document}

\title{\TN{}: A Deep Learning-Based System for Security Patch Identification
%\thanks{$^{\dagger}$The first two authors contributed equally to this work.}
%\thanks{$^*$Corresponding author: Pengbin Feng}
}

%\author{\IEEEauthorblockN{Xinda Wang, Shu Wang, Pengbin Feng, Kun Sun, Sushil Jajodia}
%\IEEEauthorblockA{\textit{Center for Secure Information Systems, George Mason University, Fairfax, VA, USA}
%^2\textit{School of Cyber Engineering, Xidian University, Shaanxi, China}
%}
%\{xwang44, swang47, pfeng4, ksun3, jajodia\}@gmu.edu}

\author{
  \IEEEauthorblockN{
    Xinda Wang\IEEEauthorrefmark{1},
    Shu Wang\IEEEauthorrefmark{1},
    Pengbin Feng\IEEEauthorrefmark{1},
    Kun Sun\IEEEauthorrefmark{1},
    Sushil Jajodia\IEEEauthorrefmark{1},
    Sanae Benchaaboun\IEEEauthorrefmark{2},
    and Frank Geck\IEEEauthorrefmark{2}
  }
  \IEEEauthorblockA{
    \IEEEauthorrefmark{1}Center for Secure Information Systems, George Mason University, Fairfax, VA, USA\\     
    \IEEEauthorrefmark{2} CSIA Division, C5ISR Center, Space and Terrestrial Communications Directorate, \\ U.S. Army Combat Capabilities Development Command (DEVCOM)\\
  }
  \IEEEauthorblockA{
    \{xwang44, swang47, pfeng4, ksun3, jajodia\}@gmu.edu, 
    \{sanae.m.benchaaboun.civ, frank.c.geck.civ\}@mail.mil}
 
% {\phantom{.}}
}

% \and
% \and
% \and
% \and
% \and

\maketitle

\begin{abstract}

%With the increasing usage of open source software (OSS) components, vulnerabilities embedded in them are propagated to a huge number of underlying applications. Although timely applying security patches is an effective defense, our prior work shows there exists a significant delay for downstream software. The main reason is that such patches do not report to CVE or even explicitly indicate their security impacts in the metadata, which is hard to be recognized by maintainers. In contrast, armored attackers can still identify these “secret” security patches and generate corresponding exploits. Therefore, it is critical to identify the existence of security patches to enable timely fixes.

%To identify the security patches, we develop a recurrent neural network (RNN) based toolkit called PatchRNN. To test its performance in the real world, we conduct a case study on an open-source web server software - NGINX. The results show that our toolkit can successfully detect secret security patches with a low false positive rate.

With the increasing usage of open-source software (OSS) components, vulnerabilities embedded within them are propagated to a huge number of underlying applications. In practice, the timely application of security patches in downstream software is challenging. The main reason is that such patches do not explicitly indicate their security impacts in the documentation, which would be difficult to recognize for software maintainers and users. However, attackers can still identify these ``secret" security patches by analyzing the source code and generating corresponding exploits to compromise not only unpatched versions of the current software, but also other similar software packages that may contain the same vulnerability due to code cloning or similar design/implementation logic. Therefore, it is critical to identify these secret security patches to enable timely fixes. To this end, we propose a deep learning-based defense system called \TN{} to automatically identify secret security patches in OSS. Besides considering descriptive keywords in the commit message (i.e., at the text level), we leverage both syntactic and semantic features at the source-code level. To evaluate the performance of our system, we apply it on a large-scale real-world patch dataset and conduct a case study on a popular open-source web server software - NGINX. Experimental results show that the \TN{} can successfully detect secret security patches with a low false positive rate.

%This work could help boost vulnerability repair efficiency and code resilience to build a more reliable and secure information system infrastructure.

\end{abstract}

\begin{IEEEkeywords}
security patch, open source software, deep learning, vulnerability mitigation
\end{IEEEkeywords}
\section{Introduction}

With the wide adoption of Open Source Software (OSS), software maintainers are usually overwhelmed by a large number of patches including both security patches and non-security patches. 
Since applying patches would introduce system downtime and extra workload (e.g., changing code for new API), security patches should take precedence over non-security patches to be applied for avoiding the N-day attack.
However, software vendors may not provide sufficient information in the changelog when releasing these patches. 
In such a case, for maintainers, manually identifying the existence of security patches is time-consuming, labor-intensive, and error-prone. Failing to timely apply a security patch of Apache Struts 2, Equifax experienced a data breach~\cite{Equifax} and exposed millions of personal information. 
To solve these problems, we propose \TN{}, a deep learning-based system to help automatically identify the security patches.

To alleviate human efforts, machine learning has been widely adopted in patch analysis since it can automatically identify target code samples containing similar patterns with those in the training data~\cite{zhou2017automated, das2018security,goseva2018identification}. Even though, machine learning still requires human experts to define a set of distinguishable features. 
In contrast, deep learning is able to automatically extract features and learn their importance from training samples~\cite{li2018vuldeepecker}.
While neural networks have not been employed in patch analysis, their successful application in natural language processing suggests that they may be suitable for programming language processing where context is also important.
Therefore, we adopt Recurrent Neural Networks (RNNs) that are effective in processing sequential and context-sensitive data~\cite{rnn_plp}.

For an OSS project maintained using Git, a patch (i.e., a commit) is composed of two parts: commit message and source code change. 
Previous patch analysis works are mainly at the text level, i.e., analyzing security-related keywords in the commit message~\cite{bavota2016mining,santos2016judging}. However, software maintainers may not explicitly and accurately specify the security impact due to limited security expertise, different human subjectivity, and changed maintenance policy~\cite{news}. In this case, we propose to consider information provided at the source-code level. Although some approaches also make use of source code in the patch, they extract the metadata like the number of lines, hunks, etc. Instead, we focus on the syntax and semantics of the source code itself.
%One reason is that it is challenging to parse a patch including both original and patched code and is not a complete program unit 

In our work, we use Recurrent Neural Network (RNN) to extract the syntax and semantic level information from patches. We utilize the information from both the diff code and commit message to capture more comprehensive features.
For the commit message, we use a TextRNN model to get the message vector. For the diff code, we use a twin RNN-based network with two identical sub-networks to obtain a code vector. After aggregating these two vectors, we can have the final prediction results by adopting a two-layer fully connected network. Our experimental results on a large-scale real-world patch dataset show that we can achieve a total accuracy of 83.57\% with an F1 score of 0.747. At the same time, the fall-out rate (false positive rate) is 11.58\% while the miss rate (false negative rate) is 26.34\%.

To further evaluate the effectiveness of our system, we perform a case study on a popular open-source web server software - NGINX, and discover 10 commits that make security-related changes but are not explicitly described in the official changelog. Among them, corresponding vulnerabilities of 5 patches are ranked as top dangerous software weaknesses~\cite{CWE}.

% contributions
\section{Background}

This section provides the definition and composition of software patches. We also describe the differences between security and non-security patches.

\lstdefinestyle{lst}{
    float=tp,
    floatplacement=tbp,
    %abovecaptionskip=0.01in,
    numbers=left, 
    numberstyle=\scriptsize, 
    numbersep = 5pt,
    framexleftmargin = 0in,
    framexrightmargin = 0in,
    xleftmargin = 0.18in,
    xrightmargin = 0.1in,
    basicstyle=\ttfamily\scriptsize, 
    frame=lines,
    showtabs=true,
    showspaces=true,
    showstringspaces=false,
    literate={\ }{{\ }}1,
}

\lstset{belowskip=-0.05in}

\begin{lstlisting}[
language=diff, 
style=lst,
caption=An example of security patch for NULL pointer dereference vulnerability (CVE-2018-19200).,
label=code1,
]
From f58c25069cf4a986fe17a80c5b38687e31feb539 Mon Sep 17 00:00:00 2001
From: Sebastian Pipping <sebastian@pipping.org>
Date: Wed, 10 Oct 2018 14:49:51 +0200
    
    ResetUri: Protect against NULL

diff --git a/src/UriCommon.c b/src/UriCommon.c
index 3775306..039beda 100644
--- a/src/UriCommon.c
+++ b/src/UriCommon.c
@@ -75,6 +75,9 @@
 
void URI_FUNC(ResetUri)(URI_TYPE(Uri) * uri) {
+   if (uri == NULL) {
+       return;
+   }
    memset(uri, 0, sizeof(URI_TYPE(Uri)));
 }
 }
\end{lstlisting}

\begin{lstlisting}[
language=diff, 
style=lst,
caption=An example of non-security patch in \textit{GoAhead} software.,
label=code2,
]
commit ac367d7a2884aa150cdfc0495348fd886d3bd228
Author: Embedthis Software <dev@embedthis.com>
Date:   Thu Nov 12 10:59:07 2015 -0800

    FIX: don't try to catch SIGKILL

diff --git a/src/goahead.c b/src/goahead.c
index 6e6c806a..aa66d292 100644
--- a/src/goahead.c
+++ b/src/goahead.c
@@ -204,7 +204,6 @@ static void initPlatform()
 {
 #if ME_UNIX_LIKE
     signal(SIGTERM, sigHandler);
-    signal(SIGKILL, sigHandler);
     #ifdef SIGPIPE
         signal(SIGPIPE, SIG_IGN);
     #endif
\end{lstlisting}

\vspace{1.25mm}
\noindent\textbf{Software patch.} A software patch is a set of code changes between two versions to address security vulnerabilities, resolve functionality bugs, or add new features. 
On a version control platform like GitHub~\cite{GitHub}, a commit can be regarded as a patch since it is good practice to separate updates for different issues. As shown in Listing~\ref{code1} and~\ref{code2}, each commit is mainly composed of two parts: the commit message that describes the commit using the natural language and the source code difference. A set of consecutive removed and added statements (i.e., lines start with \texttt{-} or \texttt{+}) with their context lines is called one hunk.
Besides, a 20-byte long hash string is used to uniquely identify a patch. A patch may have more than one hunk that modifies multiple files and multiple functions. A line starting with \textit{diff} -{}-\textit{git} is used to point out the modified files.

\vspace{1.25mm}
\noindent\textbf{Security and non-security patch.} 
Security patches correct specific weaknesses described by a vulnerability. Non-security patches include bug fix patches and new feature patches. The bug fix patches make the software run more smoothly and reduce the likelihood of a crash by correcting the software bugs. The new feature patches add new or update existing functionality to the software.

Listing~\ref{code1} is a security patch for vulnerability CVE-2018-19200 that prevents a NULL pointer dereference by adding a NULL check and corresponding handling (Line 15, 16, and 17) for \texttt{uri} whose pointed memory would be initialized as $0$ (Line 18). Listing~\ref{code2} shows an example of non-security patch. Since SIGKILL will cause the target process to terminate immediately, it cannot be intercepted. Thus, this patch avoids catching SIGKILL by removing the corresponding function call (Line 15).

\section{Methodology}

\subsection{Overview}

\begin{figure*}[t]
\centerline{\includegraphics[width=0.9\linewidth]{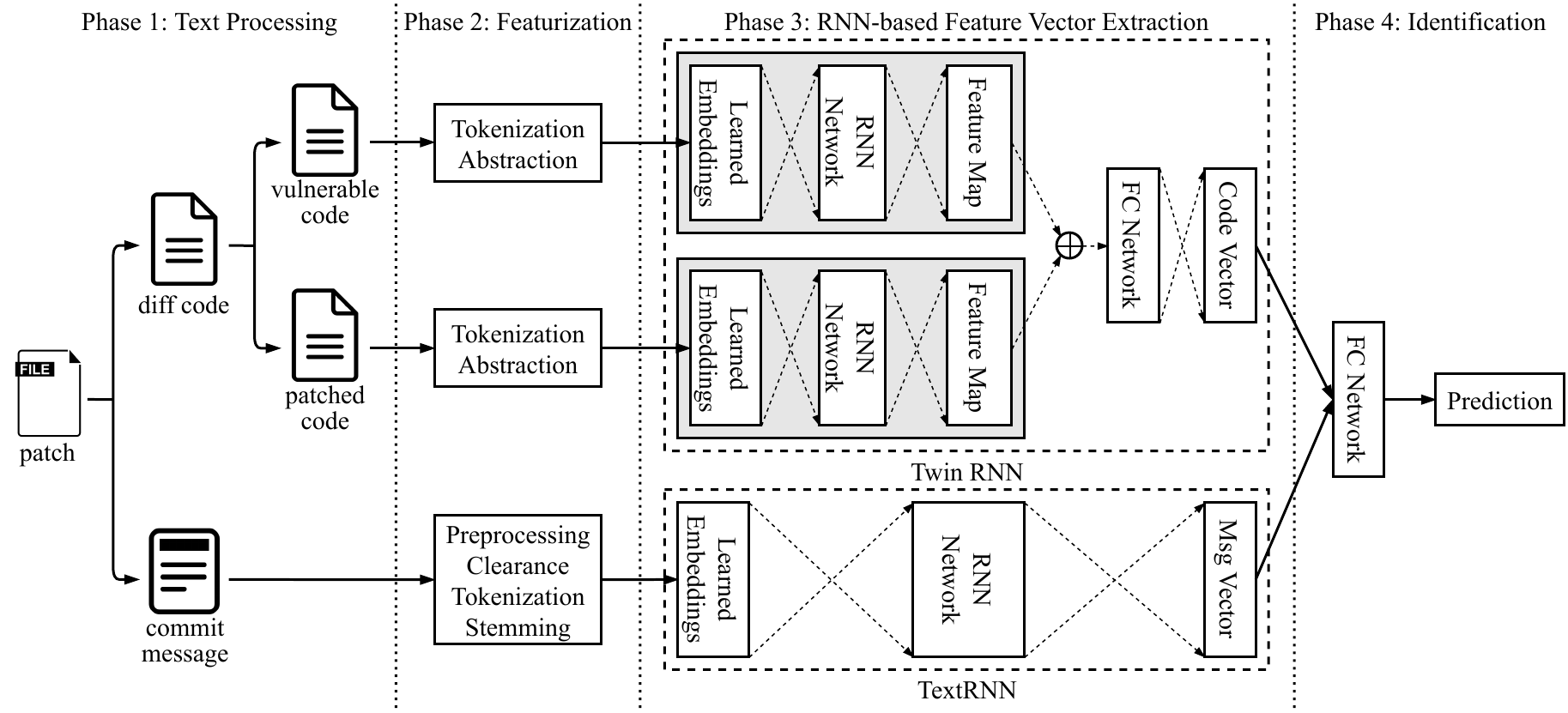}}
\centering
\centering\caption{The architecture of PatchRNN model.}
\centering
\label{fig:arch}
\end{figure*}

Figure~\ref{fig:arch} presents the architecture of our PatchRNN toolkit. 
Since a commit (i.e., a patch file) is composed of source code and commit message, we utilize both parts to capture more comprehensive features. 
% diff code
For diff source code, we reconstruct the unpatched code and patched code and process them separately with the same tokenization and abstraction strategies.
Then a twin RNN model is adopted to generate the code vector representation.
% commit msg
For the commit message, we utilize the NLP toolkit to process the text sequences and employ a TextRNN model to obtain vector representation for the commit message.
The final results are derived from the feature vectors of both parts.
We use a large-scale patch dataset \textit{PatchDB}~\cite{patchDB} to train the model. The size of the dataset is 38K.

\subsection{Feature Extraction}

\begin{figure}[h]
    \centering
    \includegraphics[width=0.9\linewidth]{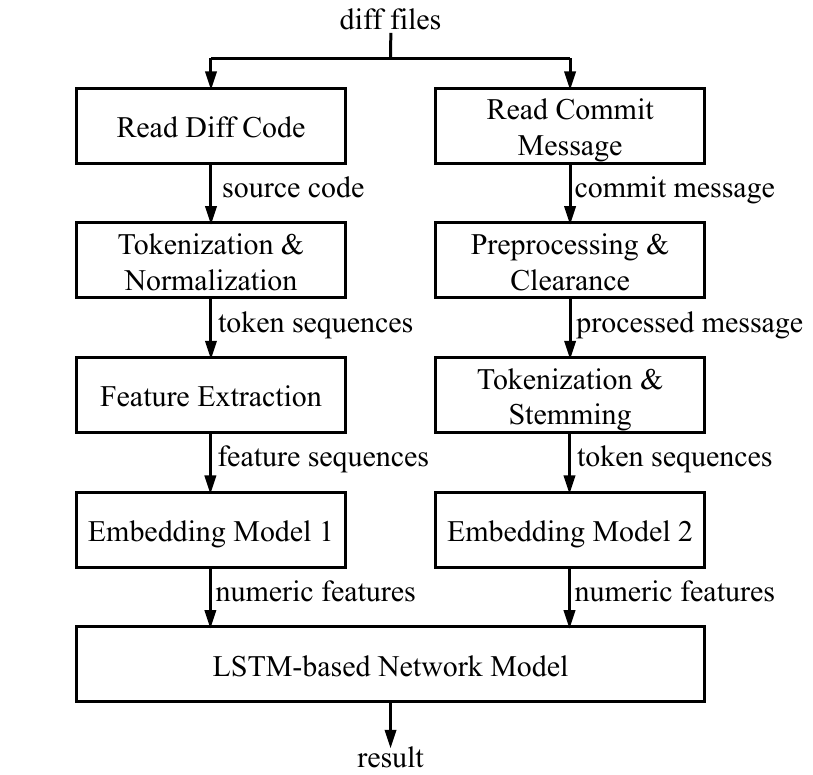}
    \caption{The overview of patch identification scheme.}
    \label{fig:features}
\end{figure}

To identify the security patches, we focus on the syntax-level information in patches, which can be learned directly by recurrent neural networks.
The overview of the patch identification scheme is shown in Fig. \ref{fig:features}.
In our scheme, the patch feature extraction contains two parts: diff code processing and commit message processing. 
We utilize the information from both the diff code and commit message to capture more comprehensive features. 
A diff file not only contains the diff code that can indicate the differences between unpatched code and patched code but also contains the commit message that may contain some clues to indicate if this patch is security-related.
As a result, the commit message is also a critical component in our scheme.

\subsubsection{Feature Extraction from Diff Code}

First, we extract the diff source code from diff files and reconstruct the original unpatched code and patched code. 
Each source code will be concatenated into a sequence and will be separated into code tokens with the \emph{Clang} tool~\cite{Clang}. 
The tokenization with Clang is a critical step since these tokens can become the direct inputs to a deep learning model. 
We find that all the tokenization tools in natural language processing (NLP) cannot work well for the programming language. 
For instance, the `equal to' operator (`==') will be separated as two `equal' signs (`=', `='), which do not conform to the meaning of programs. 
Instead, Clang is developed for C/C++ language thus is suitable for code tokenization. 
Moreover, for each token, we utilize Clang to extract additional features (e.g., the token type), which can better help us identify the security patches with syntax information. 
Except the special pad token (i.e., `\emph{$<$pad$>$}'), we have five token types for normal tokens (i.e., \emph{TokenKind.Keyword}, \emph{TokenKind.Identifier}, \emph{TokenKind.Literal}, \emph{TokenKind.Punctuation}, and \emph{TokenKind.Comment}). 
Another important feature is the diff type, which is a numeric feature that indicates if a token appears in a deleted line ($-1$), added line ($+1$), or a contextual line ($0$). 
In our design, the diff type feature is suitable for the patch file to distinguish the vulnerable code and patched code.

\begin{figure*}[h]
\centering
    \subfloat[original diff code.]{\includegraphics[width=0.505\linewidth]{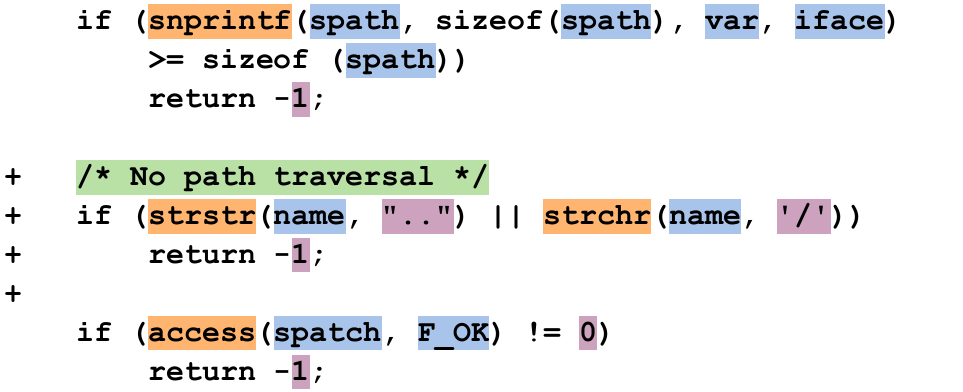}}% 
    \subfloat[abstracted diff code.]{\includegraphics[width=0.505\linewidth]{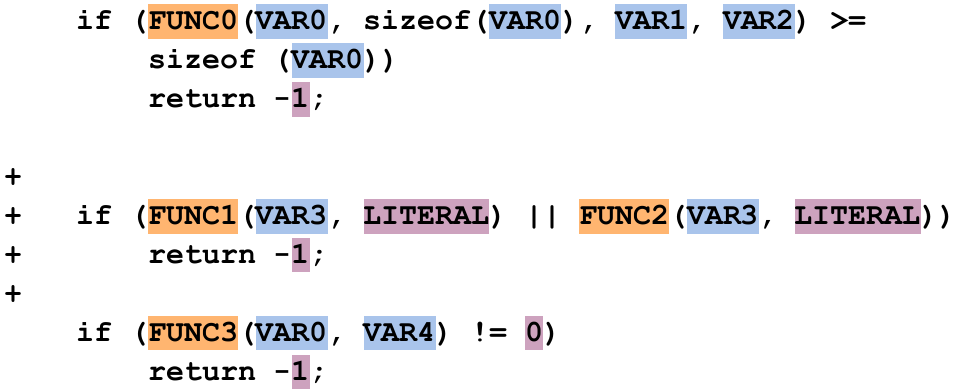}}% 
    \caption{The token abstraction processing of diff code.}
    \label{fig:abstraction}
\end{figure*}

Second, we abstract the code tokens according to their token types.
The purpose of abstraction is to reduce the token variants and vocabulary size so as to achieve a better generalization ability of the deep learning model. 
As we know, some code tokens (e.g., variable names and function names) are defined by programmers and thus have a lot of freedom in naming. 
It is inappropriate to assign each token with an individual embedding vector. 
Therefore, to handle the overfitting problem, we need to group the tokens with similar properties into one class.
After the token abstraction, the vocabulary size of the dataset is reduced from over 600K to 28K.
An example of token abstraction is illustrated in Fig. \ref{fig:abstraction}. 
We will remain the keywords and punctuation unchanged and only abstract the tokens of identifiers, literals, and comments. 
For the tokens of identifiers (i.e., variable names and function names), the abstracted tokens would be \emph{VARn} and \emph{FUNCn}, respectively. 
For the tokens of literals, we further divide the literals into constants and strings. 
We only abstract the strings into a fixed token \emph{LITERAL} but leave the constants unchanged, because some security issues (e.g., buffer overflow) are highly related to the index. 
For the tokens of comments, we delete them because the comments make little contribution to our patch identification tasks.

Third, we normalize the token sequences to a fixed length. 
We analyze the cumulative distribution function (CDF) of the token sequence length and find most of the token sequences have a short length. 
We set the normalized sequence length as 1,100 tokens because it can cover 95\% of patch samples. 
If the sequence length is less than 1,100 tokens, we will pad a special token `\emph{$<$pad$>$}' at the end of the sequence.

Finally, we convert each code token into an embedding vector by adopting an embedding model \emph{word2vec} that utilizes the skip-gram and CBOW algorithms.
The embedding vector is a learned dense representation for token features where tokens of the same meaning will have a similar representation. 
In our design, the dimension of code embeddings is set to 128.
All the features above are the inputs of the twin RNN model.

\begin{figure}[h]
    \centering
    \includegraphics[width=0.85\linewidth]{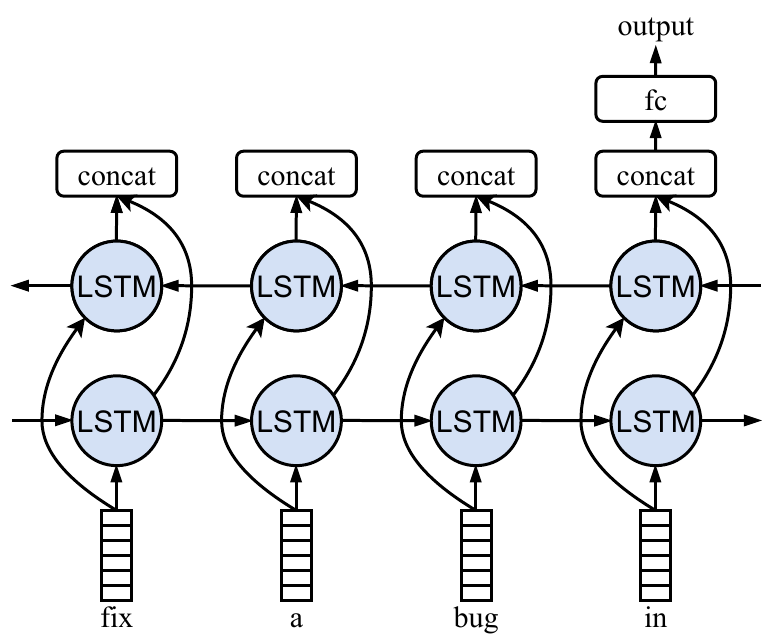}
    \caption{The structure of recurrent neural network model.}
    \label{fig:rnn}
\end{figure}

\subsubsection{Feature Extraction from Commit Message}

The commit message in a patch contains modification information such as the reason for changing the code.
For the commit message, we leverage a traditional natural language processing toolkit \emph{NLTK 3.3} for text processing, including preprocessing, clearance, tokenization, and stemming.

% preprocessing
First, all the letters in a commit message will be converted to lowercase.
% clearance.
Second, we will clear the message with regular expressions to remove unimportant information, such as URL links, independent numbers, and the signatures in the footnote.
% tokenization
Third, the processed commit message will be separated into a set of word tokens with a tokenization toolkit \emph{TweetTokenizer}. 
We further clear the tokens without any English letter and the tokens of the email address.
We also remove the tokens in the English \emph{stopwords} list since these words cannot provide any unique information for security patch identification.
% stemming
Finally, a stemming tool \emph{PorterStemmer} is utilized to stem the word tokens, reducing the inflected words or derived words to their base form.
The reason that we stem the word tokens is to improve the generalization performance of the learned model and reduce the vocabulary size. 
% normalize
The sequence of word tokens will be normalized to the length of 200 words.

All the word tokens in the commit message are converted into 128-dimensional word embeddings via \emph{word2vec} embedding model.
After converting a commit message into a sequence of embedding vectors, we will feed it into a TextRNN model to obtain a message feature vector.

\subsection{Model Learning}

We use an end-to-end deep learning model to convert a patch input to a prediction label. 
The learned model is based on RNNs that are suitable for sequential data.
Given a patch, we extract the features from the diff code part and commit message part separately and feed them into the model. 
Our model contains 4 parts: twin RNN, TextRNN, feature fusion, final prediction. 

For the diff code part, we use a twin RNN-based model to obtain the code vector.
The inputs of the twin RNN are two sequences of feature vectors from unpatched code and patched code respectively.
Each feature vector in the sequences is 135-dimensional, containing a token embedding, token type, and diff type.
The token embedding is a 128-dimensional vector derived from an embedding layer.
Token type feature is a 6-dimensional one-hot vector that indicates 5 Clang-defined token types and `$<$pad$>$'.
The diff type feature is a numeric value that can be $\pm$1 or $0$.
In the twin RNN model, there are two LSTM-based sub-networks that share the same weights as well as the same functionality. 
That means the weights in the sub-networks are adjusted synchronously during model training.
Each sub-network contains 2 bi-directional LSTM layers with a hidden layer size of 32.
The outputs of two sub-networks will be concatenated and fed into a fully connected network with the dimension of [256, 128, 64]. 
Therefore, the output code vector is 64-dimensional.

For the commit message part, we use a TextRNN model to get the message vector. 
The structure of the RNN model is illustrated in Figure \ref{fig:rnn}.
A TextRNN model contains an embedding layer of 128 dimensions, a bi-directional LSTM layer of size 32.
The LSTM outputs in the last position are concatenated as the input of a fully connected network with the dimension of [64, 64].
The message vector is hence a 64-dimensional vector.

Finally, the code vector and message vector are concatenated and fed into the prediction model, which contains a 3-layer fully connected network with the dimension of [128, 32, 2] and a softmax layer.

\section{Evaluation}

\vspace{1.25mm}
\noindent\textbf{Dataset.} 
We select \textit{PatchDB}~\cite{patchDB} as the dataset in our experiments, which is a large-scale patch dataset including both security and non-security patches in C/C++ extracted from the NVD and popular GitHub repositories. 
In total, there are 38,041 patch samples composed of 12,476 security patches and 25,565 non-security ones.
Among them, we randomly choose 80\% instances as the training set and the remaining 20\% as the testing set.

\vspace{1mm}
\noindent\textbf{Setup.}
Our program is implemented using Python 3.7, while the neural network model is designed based on PyTorch 1.6. 
Our model is carried out in the Ubuntu 20.04.1 LTS environment running in Intel Xeon Gold 5122, 3.60-GHz CPU with 64-GB RAM.
We realize the neural network training by employing a CUDA-based parallel computing platform with 2 NVIDIA RTX 2080 Ti GPUs of 11 GB memory.
The batch size of the model training is set to be $512$, while the learning rate is set to be $5 \times 10^{-4}$.

\begin{table}[h]
\begin{center}
    \begin{tabular}{c|cc}
        \toprule
        ~ & Actual Non-Security & Actual Security \\
        \midrule
        Predicted Non-Security & 4515 (T.N.) & 659  (F.N.) \\
        Predicted Security & 591  (F.P.) & 1843 (T.P.) \\
        \bottomrule
    \end{tabular}
    \caption{Confusion matrix of the predictions in our model.}
    \label{tab:exp}
\end{center}
\vspace{-0.1in}
\end{table}

\vspace{1mm}
\noindent\textbf{Performance.}
In the experiments, 30K samples are used for training, 7,600 samples are used for testing.
Our confusion matrix is shown in TABLE \ref{tab:exp}.
The total test accuracy of our model is 83.57\% with the precision of 75.72\% and the recall of 73.66\%.
The F1 score is 0.747. 
The fall-out rate (false positive rate) is 11.58\% while the miss rate (false negative rate) is 26.34\%.

\vspace{1mm}
\noindent\textbf{Overhead.}
The time of preprocessing each sample in the dataset is 4.4 sec. 
For the neural network model, the training time is 27 min for 1,000 epochs.
The training phase takes up 10.2GB of GPU memory.
For a single patch, the prediction time is 5.6 sec on average including the preprocessing and resource loading.
\section{Case Study: NGINX}

%\linespread{1.15}
\begin{table}[!tb]
\begin{center}
\caption{the comparison between the number of security issues listed in the documentation and the number of security patches committed on the GitHub repository.}
  \label{tab:overview}
\begin{tabular}{c|c|c|c|c|c}
\toprule
\multirow{2}{*}{Changes w/}    & \multicolumn{3}{c|}{Documentation (\# Items)}                                                                  & \multicolumn{2}{c}{Ground Truth (\# Commits)}                              \\ \cline{2-6} 
                                 & Security & Bugfix & Other & Security & Non-Security \\ \midrule
1.19.1 & 0                                  & 6                                & 5                                        & 8                                  & 11                                     \\ %\hline
1.19.2 & 0                                  & 7                                & 2                                        & 8                                  & 7                                      \\ %\hline
1.19.3 & 0                                  & 5                                & 4                                        & 7                                  & 12                                     \\ \bottomrule

\end{tabular}
\end{center}
\end{table}

To further evaluate the performance of the \TN{}, we conduct a case study on a popular open-source web server software - NGINX and find some secret security patches.

Although maintained on GitHub~\cite{nginx}, the NGINX website~\cite{nginx_changes} would also list the changes when a new version is released. These changes would be tagged as ``security", ``bugfix", ``feature", ``change", etc. However, we find that, even though no changes are labeled as security ones for some releases, some GitHub commits are for security fixes. Such commits can be regarded as secret security patches since vendors do not explicitly release their potential security impacts, which are highly likely to be ignored by software maintainers or users. 

To figure out the number of security patches in the NGINX, we apply our tool on its GitHub commits. Since we need to manually check each commit to get the ground truth, we focus on three new versions (i.e., NGINX 1.19.1, 1.19.2, and 1.19.3).

Table~\ref{tab:overview} shows the number of security, bug fix, and other changes with NGINX 1.19.1, 1.19.2, and 1.19.3. The official documentation does not mention any security changes in these versions. However, by manually checking the GitHub commits of these versions, we find there are 8, 8, and 7 commits that are security-related, respectively. 

After applying our \TN{} on commits of the above three versions, we summarize the detection results in Table~\ref{tab:results}. In total, our toolkit identifies 10 security patches that are secretly released by NGINX.
Among all security patches, 43\% (10 out of 23) are successfully identified.
By manually checking their corresponding vulnerability types, we find 5 out of them are ranked as one of the CWE top 25 most dangerous software weaknesses~\cite{CWE}, e.g., use-after-free, NULL pointer dereference, out-of-bound access, etc.
Meanwhile, the \TN{} does not introduce any false positive cases. That is to say, no non-security patches are labeled as security-related.
In such cases, once a security patch is identified, it can be directly prioritized to be applied and it is highly likely to be a real security patch. 

%The list of secret security patches can be found in the appendix.

% \usepackage{graphicx}
\begin{table}[!tb]
\begin{center}
\caption{Detection Results on NGINX.}
  \label{tab:results}
\begin{tabular}{c|c|c|c|c|c|c}
\toprule
\multirow{2}{*}{Changes w/}      & \multicolumn{3}{c|}{Documentation} & Ground Truth & \multicolumn{2}{c}{Detection Results}                          \\ \cline{2-7} 
                                 & \multicolumn{3}{c|}{Security Issues}      & Security Patches     & { } T.P. { } & F.P. \\ \midrule
1.19.1 & \multicolumn{3}{c|}{0}             & 8                                      & 4                              & 0                              \\  
1.19.2 & \multicolumn{3}{c|}{0}             & 8                                      & 3                              & 0                              \\  
1.19.3 & \multicolumn{3}{c|}{0}             & 7                                      & 3                              & 0                              \\  \midrule
Sum.   & \multicolumn{3}{c|}{0}             & 23                                     & 10                             & 0                              \\ \bottomrule
\end{tabular}
\end{center}
\vspace{-0.15in}
\end{table}

\section{Related Works and Discussion}

\noindent \textbf{Patch Analysis.} Most of the previous works on patch analysis focus on leveraging natural language description~\cite{zhou2017automated, das2018security,goseva2018identification}. They extract textual information like keywords in the bug report, commit message, changelog, etc. However, these methods require consistent and well-maintained documentation~\cite{pereira2019identifying}, which performs limited generalization capability in newly released patches or patches for other projects.
Some works make a step forward by retrieving and analyzing the source code part of the patch~\cite{zhong2015empirical,soto2016deeper,perl2015vccfinder,machiry2020spider, tian2012identifying,wang2019detecting,wang2020machine}. They study the attributes of metadata and syntax like the number of deleted lines, conditional statements, program inputs, etc.
However, these works do not distinguish between security patches and non-security bug fixes.
Although some works~\cite{zaman2011security, li2017large} perform empirical studies on differences between security and non-security patches, they do not provide a practical method to help automatically identify security patches.

At the binary level, Xu et al. \cite{xu2017spain} analyze the difference in execution traces to identify the existence of a security patch. 
Given the source code of a security patch, some researchers~\cite{zhang2018precise, jiang2020pdiff} propose methods to test if the corresponding vulnerabilities have been patched in the binaries.

\vspace{1.25mm}
\noindent\textbf{Discussion.} %Our work has some limitations. 
Currently, the \TN{} only supports C/C++ that are languages with the highest number of vulnerabilities~\cite{white-source}. Future work can extend it by applying other analysis tools to parse other programming languages. Also, the input of our system is not limited to GitHub commits. For projects that are not maintained by Git, its changelog can be used as the commit message, and code difference can be computed between two neighboring versions.

\section{Conclusion}

In this work, we propose the first deep learning-based approach that automatically identifies security patches to prioritize their application. We leverage both the commit message and source code difference to overcome the situation where the documentation of a patch is not well-maintained. Then, we apply TextRNN and twin RNN, respectively. The evaluation results on a real-world large-scale patch dataset and well-known web server software show that we could achieve good performance with low false alarms. 
\section{Acknowledgments}

This work was partially supported by the US Department of the Army grant W56KGU-20-C-0008 and the National Science Foundation grant CNS-1822094.

%-------------------------------------------------------------------------------
\bibliographystyle{plain}
\bibliography{reference}

\end{document}